\documentclass[lettersize,journal]{IEEEtran}
\usepackage[dvipdfm]{graphicx} % Required for inserting images
\usepackage{subcaption}
\usepackage{amsmath, amssymb, amsfonts, amsthm, empheq, cases}
\usepackage[colorlinks=true]{hyperref}
\usepackage{xcolor}
\usepackage{enumitem}
\usepackage{subcaption}

\theoremstyle{definition}

\newtheorem{theorem}{Theorem}

\newtheorem{remark}{Remark}

\newcommand\RR{\mathbb{R}}
\newcommand\sgn{\mathrm{sgn}}

\newcommand\Interpreter{\mathcal{A}}
\newcommand\Plant{\mathcal{P}}

\newcommand\Controller{\mathcal{C}_\theta}

\newcommand\IntentExtractor{f_\mathrm{int}}
\newcommand\ParameterUpdater{f_\mathrm{up}}

\newcommand\StepInterval{\Delta t}

\newcommand{\tcb}[1]{#1}%{\textcolor{blue}{#1}}
\newcommand{\tcc}[1]{#1}%{\textcolor{magenta}{#1}}

\newtheorem{assumption}{Assumption}
\newtheorem{corollary}{Corollary}
\title{Chat-Driven Reconfiguration of Model Predictive Control}
\author{
Yuya Miyaoka, Masaki Inoue, \IEEEmembership{Member, IEEE}, Jos{\'e} M Maestre, 
\IEEEmembership{Senior Member, IEEE}
\thanks{This work was supported by Grant-in-Aid for Scientific Research (B), No.~20H02173 and 25K01254 from JSPS.}
\thanks{Y. Miyaoka and M. Inoue are with the Department of Applied Physics and Physico-Informatics, Keio University, 
Yokohama 223-8522, Japan (e-mail:miyaoka.yuya@keio.jp, minoue@appi.keio.ac.jp). 
Jos{\'e} M Maestre is with the Department of Systems and Automation
 Engineering, University of Seville, 41004 Seville, Spain
 (e-mail: pepemaestre@us.es).
}}
\date{}

\begin{document}

\maketitle

\begin{abstract}
\tcb{Traditional control personalization requires users to understand optimization parameters and provide repetitive numerical feedback, creating significant barriers for non-expert users. To deal with this issue, we propose ChatMPC, a model predictive control framework that enables users to personalize control systems and adapt to environmental changes through natural language interaction. The framework operates in two modes: personalization, where users iteratively adjust control behavior to their preferences, and co-development, where users provide real-time environmental information that complements sensor data. We establish convergence guarantees under different user behavior models, demonstrating exponential convergence for consistent feedback and finite-time convergence with logarithmic interaction complexity for tolerance-based users. We validate ChatMPC through experiments in robot navigation with personalized obstacle avoidance and semi-autonomous driving with conversational obstacle reporting. Both experiments achieve real-time performance and demonstrate effective adaptation to user preferences and environmental changes.}
\end{abstract}

% > Main/Introduction.tex >
\section{Introduction}

\tcb{
Modern control systems are increasingly operating in complex, dynamic, and uncertain environments. 
Despite notable advances in autonomy, adaptive control, and learning-based methods, fully autonomous systems struggle to guarantee safety, robustness, or task success when exposed to unpredictable conditions or incomplete information~\cite{Moghadam17, Lindemann21, Li22ITS, Aniculăesei16}. 
For example, issues such as sensor limitations, model uncertainties, and the inability to anticipate novel disturbances can commonly lead to degraded performance~\cite{Zhu20ICCAD, Carvalho14, Marshall17}. 
As a result, human supervision remains a necessary component in many real-world applications, particularly when safety or adaptability is critical.
}

\tcb{
These limitations have motivated the development of human-machine cooperative control frameworks, not just to insert the human into the loop, but to systematically integrate human intent, preferences, and observations into the control architecture. 
Examples of these research efforts include shared control~\cite{Marcano20_SharedControl, Jiang17_SharedControl, Guo19_SharedControl}, human-in-the-loop systems~\cite{Samad23_HITL, Samad20_HITL, Sadowska23_Predictive}, and human-machine teaming~\cite{Music17_HumanMachineTeaming}. Each of these frameworks differs in the timing, role, and depth of human involvement, but they all rely to some extent on effective communication between human and machine.
}

\tcb{
In this context, natural language has gained traction as a powerful interface for human-machine interaction. 
Compared to traditional tuning or manual feedback mechanisms, natural language offers a low-burden, high-bandwidth channel for users to express task objectives, constraints, or environmental updates~\cite{Ye23TrustChatGPT, Gkournelos24CIRP, Gao24RobustLLM}. 
Indeed, recent studies demonstrate that natural language interfaces (NLIs) increase usability and trust~\cite{Benjamin24ACM, Liu17ReviewNLI}, lower the entry barrier for non-experts, and improve system flexibility in dynamic environments~\cite{Siwach23IMCET, Bucker22IROS}. 
Therefore, NLIs provide us with the means to bridge the gap between intuitive and accessible human communication and the formal control specifications mastered by experts. 
In this way, NLIs open a path toward scalable and personalized autonomy for future control systems.
}

\begin{figure}[t]
    \centering
    \includegraphics[width=1\linewidth]{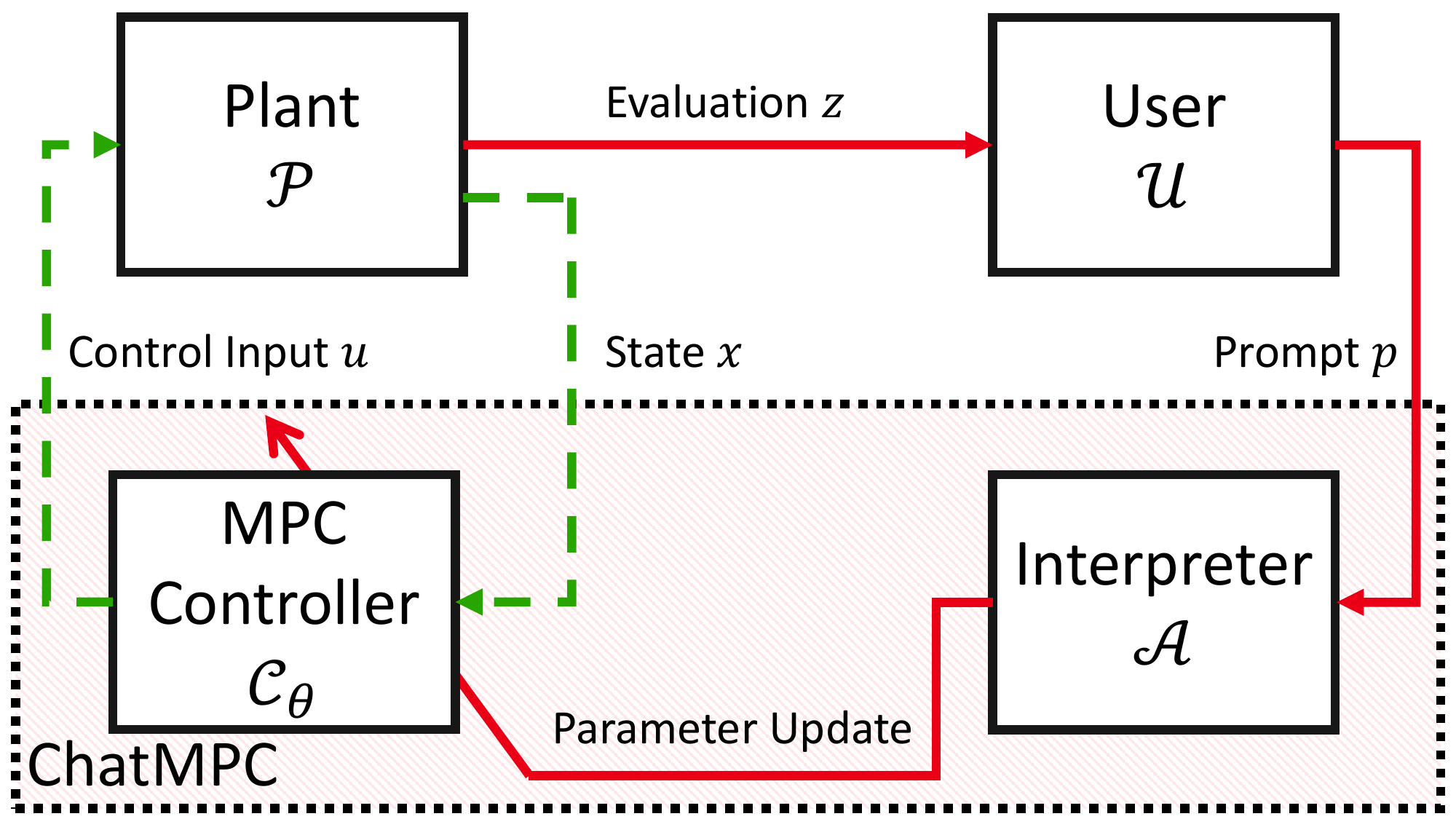}
    \caption{\tcb{
    Overview of the ChatMPC architecture. 
    In this diagram, system $\mathcal{P}$ denotes the plant, which generates an output $z$ perceived by the user, represented as system $\mathcal{U}$. 
    ChatMPC (pink-striped region enclosed by a dotted black border) is composed of an interpreter $\mathcal{A}$, which processes natural language prompts $p$ provided by the user to generate a parameter $\theta$, which is in turn fed to an MPC controller $\mathcal{C}_\theta$, where the subscript stresses that $\theta$ sets the behavior of the controller. 
    In this way, the controller $\mathcal{C}_\theta$ can generate the control input $u$ based on the current state of the plant $x$, benefiting from the standard feedback loop (green dashed line) and personalization loop (red line), which collects the preferences of the user.}}
    \label{F:M.StructureOverview}
\end{figure}

\tcb{
In this article, we propose \textit{ChatMPC} (see Figure~\ref{F:M.StructureOverview}), a model predictive control (MPC) framework that leverages natural language input to personalize control specifications and adapt to dynamic environments \cite{EFCBOOK}. 
Unlike prior works such as InstructMPC~\cite{Wu25_InstructMPC} and LanguageMPC~\cite{Sha25_LanguageMPC}, which exploit large language models and employ static inference stages without convergence guarantees, ChatMPC allows dynamic human interaction during operation and provides formal assurance of convergence to user intent. 
Furthermore, the framework enhances situational awareness without increasing cognitive load---a common challenge in personalized control~\cite{Turco24SharedControl, Jin19MusicRec}--- because it combines user-provided information with onboard sensor data. Besides the convergence analysis of user-driven updates, this work also demonstrates two core capabilities (i) personalization of control behavior based on user preferences, and (ii) real-time environmental adaptation via natural language sensing in two  experiments. 
Finally, note that a preliminary version of this idea appeared in ~\cite{Miyaoka24_ChatMPC}, but it was based on a simpler architecture and lacked  the previously mentioned formal and theoretical aspects that are fully developed here. }

% 各ループの時間スケールの違いについての注意喚起

\tcb{
The remainder of the paper is organized as follows. 
Section~\ref{Section:M} introduces the architecture of ChatMPC. 
Section~\ref{Section:TheoreticalAnalysis} presents the theoretical analysis of convergence. 
Sections~\ref{Section:ExperimentCBFLLM} and~\ref{Section:ExperimentCarla} demonstrate ChatMPC in personalization and environmental adaptation tasks, respectively. 
Section~\ref{Section:Conclusion} concludes the paper.
}

% < Main/Introduction.tex <
% > Main/Methodology.tex >
\section{ChatMPC}\label{Section:M}

\tcb{As can be seen in Figure~\ref{F:M.StructureOverview},  the overall architecture of ChatMPC  integrates a conventional feedback loop with a novel personalization loop based on natural language interaction. In this loop, the user $\mathcal{U}$ observes an evaluation output $z$ from the plant $\mathcal{P}$ and provides a natural language \textit{prompt} $p$ to the interpreter $\mathcal{A}$, which updates the controller parameter $\theta$ based on the analysis of $p$. This mechanism allows the behavior of the MPC controller to adapt to the user's evolving preferences and environmental understanding. In this way, ChatMPC enables real-time adaptation and specification refinement through intuitive low-burden communication in a different time scale (e.g., in the case of an autonomous electric wheelchair, the control loop functions on a millisecond timescale, while the personalization loop updates only intermittently), distinguishing it from other MPC schemes with static or manually tuned parameters. }

% 各小節の予告
In the following subsections, we provide details of each component of ChatMPC. We describe the MPC controller in Subsection~\ref{Subsection:M.MPCController} and the interpreter in Subsection~\ref{Subsection:M.Interpreter}.

\subsection{\tcb{Plant}}
\label{Subsection:M.Plant}

% プラントの紹介
First, we consider a  plant  described by the discrete-time state-space representation:
\begin{equation}
    \Plant :
    \begin{cases}
        x(k+1) = f(x(k),u(k)),\\
        z(k) = g(x(k),u(k)),
    \end{cases}
    \label{E:M.Plant}
\end{equation} % プラントの状態方程式
where $x\in\RR^n$ and $u\in\RR^m$ are the plant state and the plant input, respectively, and $z\in\RR^l$ is the evaluation output. Symbols $f : \RR^n\times\RR^m \to \RR^n$ and $g : \RR^n\times\RR^m \to \RR^l$ represent the plant state and the evaluation output mapping, respectively.

\subsection{User}
\label{Subsection:M.User}

\tcb{The user $\mathcal{U}$ functions as a high-level, event-triggered oracle whose feedback drives the long-term adaptation of the controller. To this end, the user provides natural language prompts $p(\tau)$ in response to the observed \emph{controlled} plant behavior. These prompts encapsulate implicit preferences, concerns, or goals that may not be formally specified in the original control objectives. Examples include ``It is too close to that child'' and ``Could you hurry up, please?''.}

\tcb{Unlike the control loop, which operates at high frequency, the user interacts asynchronously and at a much slower timescale, as driven by contextual awareness and comfort. This separation allows $\mathcal{U}$ to intervene only when necessary, minimizing cognitive burden while maintaining the ability to guide the system's behavior.}

\tcb{Finally, from a system design perspective, notice that $\mathcal{U}$ is not modeled as a deterministic agent but rather as an open-ended source of qualitative feedback where the language interface acts as a bridge between human intuition and control formalism, enabling non-expert users to steer complex systems without modifying low-level parameters directly.}

\subsection{MPC Controller}
\label{Subsection:M.MPCController}
% MPCコントローラの最適化問題
\tcb{Consider a} MPC controller with a step interval of $\StepInterval$ \tcb{and} a prediction horizon of $H$. At each step $k$, the controller calculates the optimal input $u(k|k)$, solving the following optimization problem:
\begin{subequations}
\begin{empheq}[left={\Controller: \empheqlbrace}]{align}
    \min_{X(k), U(k)}~
        & J_\theta(X(k),U(k)) ,  \label{E:M.MPCOptimizationProb.Obj}\\
    \text{s.t.}~
        & x(k|k)=x(k), \\
        \begin{split}
        & x(k+i+1|k)=\\
            & \qquad f(x(k+i|k),u(k+i|k)), \\
            & \qquad \forall i \in \{0,\ldots,H-1\}, 
        \end{split}\\
        & X(k) \in \mathcal{X}_\theta, \label{E:M.MPCOptimizationProb.StateConstraints}\\
        & U(k) \in \mathcal{U}_\theta,
\end{empheq}
    \label{E:M.MPCOptimizationProb}%
\end{subequations}
where $J_\theta(X(k),U(k))$ is the cost function, $X(k)=[ x(k+1|k)^\top~\cdots~x(k+H+1|k)^\top ]^\top$ and $U(k)= [ u(k|k)^\top~\cdots~x(k+H|k)^\top ]^\top$\tcb{, with the notation $(k+i|k)$ referring to variable values for} time step $k+i$ calculated \tcb{at} time step $k$. \tcb{Here}, $\theta\in\RR^q$ is the adjustable parameter\tcb{, which defines the sets} $\mathcal{X}_\theta$ and $\mathcal{U}_\theta$ of the allowable state sequence and  input sequences \tcb{so as to adapt them to the user preferences. Note that the cost function} $J_\theta$ \tcb{is also} parameterized by $\theta$. \tcb{Therefore, the} goal of the personalization loop is to update $\theta$ to align \tcb{the behavior of the controller} with user's implicit preferences.

\subsection{Interpreter}
\label{Subsection:M.Interpreter}

\begin{figure}[t]
    \centering
    \includegraphics[width=1\linewidth]{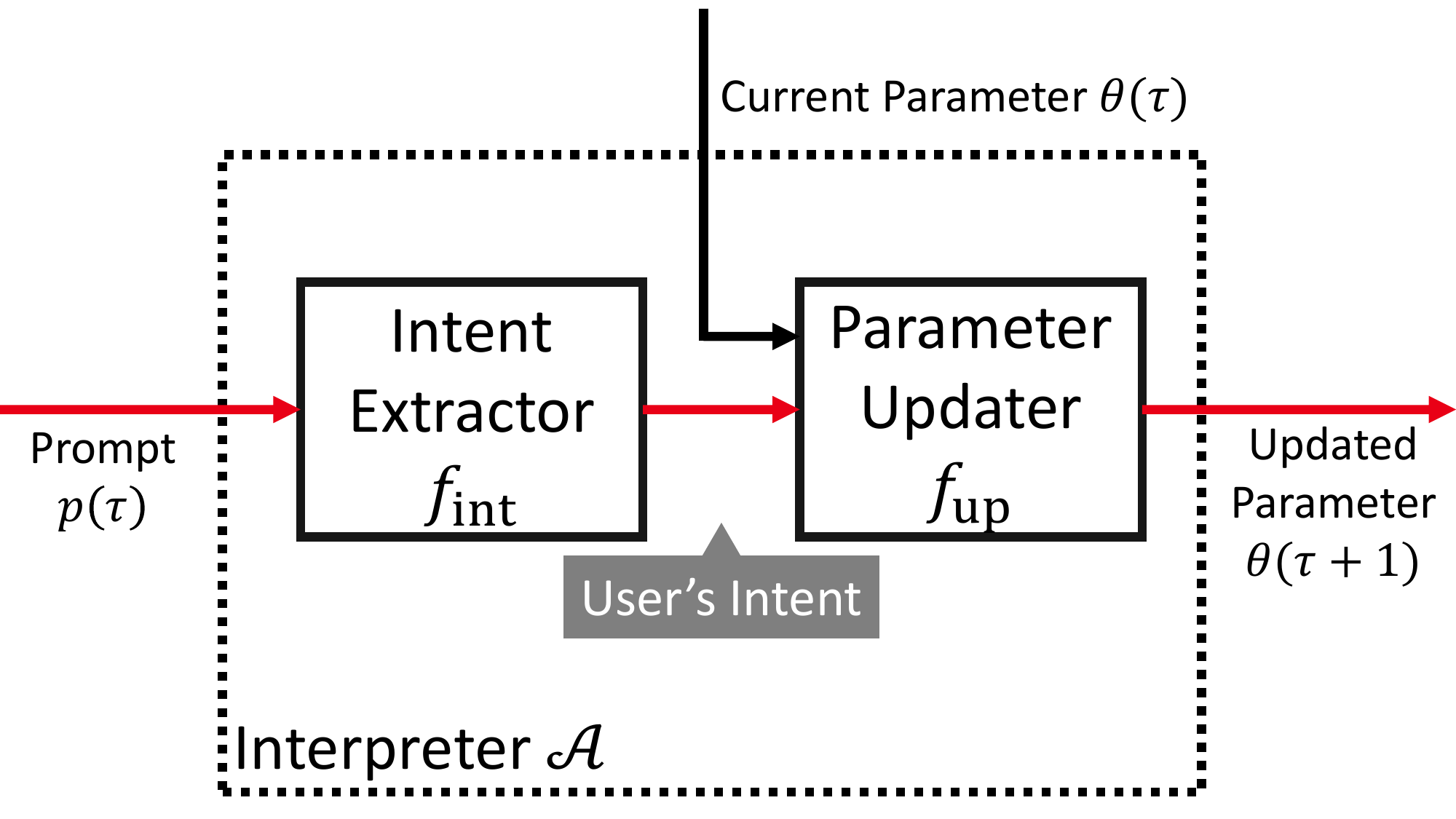}
    \caption{\tcb{Structure of the interpreter $\Interpreter$ used in ChatMPC. It receives user prompts $p(\tau)$ and processes them in two stages: (i) an intent extractor $\IntentExtractor$ maps the prompt into an update marker; (ii) a parameter updater $\ParameterUpdater$ computes the updated controller parameter $\theta(\tau+1)$.}}
    \label{F.ChatMPC.Interpreter}
\end{figure}

% インタプリタの紹介
The interpreter $\Interpreter$ is an important component of the personalization loop in ChatMPC \tcb{because it adjusts} the parameter $\theta$ according to the content of the prompts $p$ \tcb{provided by the user.} \tcb{Figure~\ref{F.ChatMPC.Interpreter} shows the structure of the interpreter $\Interpreter$,} which consists of an intent extractor $\IntentExtractor$ and a parameter updater $\ParameterUpdater$. 

% インタプリタの構造

\tcb{The execution of the interpreter $\Interpreter$ is triggered by the user prompts in an event-based fashion. To reflect this, we denote each execution by the iteration index} $\tau\in\{0,1,\ldots\}$. \tcb{Once} the interpreter $\Interpreter$ receives the $\tau$-th prompt $p(\tau),\tau\in\{0,1,\ldots\}$ from the user\tcb{, the following operations are performed:}
\begin{enumerate}
    \item The intent extractor $\IntentExtractor$ analyzes the content of the prompt $p(\tau)$ and outputs \tcb{the update marker} $s(\tau) \in\RR^q$ based on \tcb{its} intent, \tcb{providing} the information about which element of parameter $\theta(\tau)$ should be updated. \tcb{This is} expressed as follows:
\begin{equation}
    \IntentExtractor: s(\tau) = \IntentExtractor(p(\tau)) \in\{-1,0,+1\}^q
    \label{E.ChatMPC.ParameterUpdateMarker}
\end{equation} % 修正マーカー
\noindent \tcb{where} a non-zero element indicates that the corresponding element of the parameter should be updated \tcb{and the sign indicates the direction of the update}. 

For the implementation of $\IntentExtractor$, it can be effective to incorporate natural language models such as the Sentence BERT model \cite{Nils19} or few-shot learning with LLMs. \tcb{Specific details of the implementation in this article are given in the experimental Sections~\ref{Section:ExperimentCBFLLM} and \ref{Section:ExperimentCarla}}.

    \item The parameter updater $\ParameterUpdater$ calculates the updated $\theta(\tau+1)$ \tcb{in an element-wise manner} based on the update marker $s(\tau)$ and the previous parameter $\theta(\tau)$ as:
    \begin{align}
    \ParameterUpdater: 
     \theta(\tau+1) &= \theta(\tau) + s(\tau) \odot \eta(\tau), \label{E:M.ParameterUpdater}
\end{align}
\noindent \tcb{where the size of the update is governed by}
\begin{align}
    \eta(\tau)_i &= \begin{cases}
        \gamma_i \eta(\tau-1)_i ,& s(\tau)_i s(\tau-1)_i = -1 ,\\
        \eta(\tau-1)_i ,& \text{else}.
    \end{cases}
    \label{E:M.ParameterUpdater-eta}
\end{align}
\tcb{Here, $\gamma\in(0,1)^q$ imposes a decay rate that sets the evolution of $\eta(\tau), \tau\in\{1,2,\ldots\}$. The rationale for this logic is that it is preferable to decrease} the size of the update $\eta$ when the update marker \textit{flips} compared to the previous time \tcb{because this reflects that the behavior of the controller went beyond the implicit intent of the user}.

\tcb{Notice that initial values} $\theta(0) = \theta_0 \in \RR^q$ and $\eta(0) = \eta_0 \in \RR_+^q$ \tcb{must be provided by the system designer.}
\end{enumerate}

% パラメータ更新器の説明

% The first update, from $\tau=0$ to $\tau=1$, is performed as follows:
% \begin{align}
%     \theta(1) = \theta_0 + s(0) \odot \eta_0,
%     \label{E:M.ParameterUpdaterAtInitialTime}
% \end{align}
% where $\theta_0$ is the initial guess of the specification parameter, and $\eta_0\in\RR_+^q$ is a hyperparameter that determines the size of the update.
% Subsequent parameter updates are performed as follows
% %%%
% \begin{align}
%     \ParameterUpdater: 
%      \theta(\tau+1) &= \theta(\tau) + s(\tau) \odot \eta(\tau), \label{E:M.ParameterUpdater}
% \end{align}
% %%%
% where each element of $\eta(\tau)$ follows
% %%%
% \begin{align}
%     \eta(\tau)_i &= \begin{cases}
%         \gamma_i \eta(\tau-1)_i ,& s(\tau)_i s(\tau-1)_i = -1 ,\\
%         \eta(\tau-1)_i ,& \text{else}
%     \end{cases}
%     \label{E:M.ParameterUpdater-eta}
% \end{align}
% %%%
% with a decay rate $\gamma\in(0,1)^q \subset \RR_+^q$.
% \begin{subequations}
% \begin{empheq}[left={\empheqlbrace}]{align}
%     \theta(\tau+1) &= \theta(\tau) + s(\tau) \odot \eta(\tau), \label{E:M.ParameterUpdater-a}\\
%     \begin{split}
%     \eta(\tau)_i &= \begin{cases}
%         \gamma_i \eta(\tau-1)_i ,& s(\tau)_i s(\tau-1)_i = -1 ,\\
%         \eta(\tau-1)_i ,& \text{else} ,
%     \end{cases}\\
%     &\quad \forall i \in \{1,\ldots,n\},\label{E:M.ParameterUpdater-b}
%     \end{split}
% \end{empheq}\label{E:M.ParameterUpdater}%
% \end{subequations}
% where $\gamma\in(0,1)^q \subset \RR_+^q$ is a decay rate.

% 注釈 意図抽出器とパラメータ更新器の実装方法について
\begin{remark}
% Section~\ref{Section:ExperimentCBFLLM}.
% In addition, the update logic of $\ParameterUpdater$ can be different from \eqref{E:M.ParameterUpdaterAtInitialTime} and \eqref{E:M.ParameterUpdater}.
The parameter update logic is not limited to the forms presented in \eqref{E:M.ParameterUpdater} and \eqref{E:M.ParameterUpdater-eta}. For instance, if the parameter values must be positive, exponentially increasing or decreasing them allows a smooth update within that range.
Specifically, one can modify the additive-type update logic given in \eqref{E:M.ParameterUpdater} and \eqref{E:M.ParameterUpdater-eta} to the multiplicative-type one, as follows:
\begin{align*}
    \theta(\tau+1)_i=\eta(\tau)_i^{s(\tau)_i} \theta(\tau)_i.
\end{align*}
\end{remark}
% < Main/Methodology.tex <
% > Main/TheoreticalAnalysis.tex >
\section{\tcb{Convergence of ChatMPC}}\label{Section:TheoreticalAnalysis}

\tcb{This section establishes the convergence properties of the personalization loop in ChatMPC through two user models, demonstrating that ChatMPC can reliably adapt to user preferences under reasonable assumptions. }

\subsection{\tcb{Exponential Convergence Analysis}}\label{Subsection:ExponentialAnalysis}

\tcb{This analysis models the user as having an implicit convex objective function and demonstrates exponential convergence to the optimal parameter configuration.}

% 勾配に関する大きな仮定
\begin{assumption}[Gradient Oracle]\label{A:GradientOracle}
\tcb{The user has an implicit convex objective $J_{\rm h}(\theta): \mathbb{R}^q \to \mathbb{R}$ with unique minimizer $\theta^*$. By \textit{evaluating} the control performance, the user indicates the \textit{direction} of his/her preference through the $\tau$-th prompt, $p(\tau)$, and the interpreter $\IntentExtractor$ the intent extractor correctly identifies the gradient direction: $$s(\tau) = \IntentExtractor(p(\tau))= -\text{sgn}(\nabla J_h(\theta(\tau))),$$
where $\sgn(\cdot)$ is the element-wise sign function that indicates $+1$ for positive elements, $-1$ for negative elements, and $0$ for zero elements.}
\end{assumption}

% 定理
\begin{theorem}[Exponential Convergence]\label{T:ExponentialConvergence}
\tcb{Under Assumption~\ref{A:GradientOracle}, the parameter sequence $\theta(\tau)$ converges exponentially to $\theta^\ast$. In particular, for any $\theta_0$, there exist positive constants $\alpha > 0$ and $C = \lceil\gamma^{-1}\rceil$ such that:
\begin{equation}
\|\theta(\tau) - \theta^*\|_2 \leq \alpha \gamma^{\tau/C}.
\end{equation}}
\end{theorem}

% 証明
\begin{proof}
% 符号勾配の簡単な表記
\tcb{Let us analyze first  the scalar case, i.e., $q=1$, and let us define the error $e(\tau) = \theta(\tau) - \theta^*$. From Assumption~\ref{A:GradientOracle} and the convexity of $J_{\rm h}$, we have:}
\begin{equation}
s(\tau) = -\text{sgn}(\nabla J_h(\theta(\tau))) = -\text{sgn}(e(\tau)).
\label{E:SignRelation}
\end{equation}

% 折り返しの紹介
\tcb{A \emph{flip} at iteration $\tau$ occurs if $\text{sgn}(e(\tau)) \neq \text{sgn}(e(\tau-1))$, indicating that the parameter has crossed the optimal value $\theta^*$. Let $\{\tau_\mu\}_{\mu=1}^{\infty}$ denote the sequence of flip iteration indices, where $\mu-1$ counts the number of flips before $\tau_\mu$.} 

% 折り返し偏差の上限
\tcb{At each flip time $\tau_\mu$, the error exhibits geometric decay due to the parameter update dynamics governed by  \eqref{E:M.ParameterUpdater} and \eqref{E:M.ParameterUpdater-eta}. To see this, note that for the $\mu$-th flip, it holds:}
\begin{equation}
e(\tau_\mu) = e(\tau_\mu-1) - \eta(\tau_\mu-1) \text{sgn}(e(\tau_\mu-1)), \nonumber
\end{equation}
\tcb{where $\eta(\tau_\mu-1) = \gamma^{\mu-1} \eta_0$ because $\mu-1$ flips occurred prior to this iteration. Since the flips occur when the error changes sign, the error update requires either}
\begin{equation}
e(\tau_\mu) = e(\tau_\mu-1) - \gamma^{\mu-1} \eta_0\;\;\mathrm{with}\;\;e(\tau_\mu-1)>0,\; e(\tau_\mu)<0 \nonumber
\end{equation}
or
\begin{equation}
e(\tau_\mu) = e(\tau_\mu-1) + \gamma^{\mu-1} \eta_0\;\;\mathrm{with}\;\;e(\tau_\mu-1)<0, \;e(\tau_\mu)>0. \nonumber
\end{equation}

\tcb{Since the flip occurs precisely when the error changes sign due to the parameter update, the magnitude of the error just before the flip must satisfy $|e(\tau_\mu-1)| \leq \gamma^{\mu-1} \eta_0$. Consequently, the error magnitude after the flip is strictly bounded by:}
\begin{equation}
|e(\tau_\mu)| < \gamma^{\mu-1} \eta_0.
\label{E:FlipErrorBound}
\end{equation}

% 折り返し時刻間の上限
\tcb{Moreover, a next flip can occur only if the error crosses again to the opposite sign. During the interval $[\tau_\mu, \tau_{\mu+1})$, exactly $\mu$ flips have occurred, so the update size remains constant at $\eta(\tau) = \gamma^\mu \eta_0$. Since $|e(\tau_\mu)| < \gamma^{\mu-1} \eta_0$, the number of iterations required to reduce this error to zero and cross to the opposite sign is at most:}
\begin{equation}
\tau_{\mu+1} - \tau_\mu \leq \left\lceil \frac{|e(\tau_\mu)|}{\gamma^\mu \eta_0} \right\rceil < \left\lceil \frac{\gamma^{\mu-1} \eta_0}{\gamma^\mu \eta_0} \right\rceil = \lceil \gamma^{-1} \rceil =: C.
\label{E:FlipIntervalBound}
\end{equation}

% 折り返し時刻間の単調減少性
\tcb{Between consecutive flips $\tau_\mu$ and $\tau_{\mu+1}$, the sign of the error $e(\tau)$ remains constant by definition. This implies that each parameter update consistently moves $\theta(\tau)$ toward the optimal value $\theta^*$, resulting in:}
\begin{equation}
|e(\tau)| \leq |e(\tau_\mu)| < \gamma^{\mu-1} \eta_0, \quad \forall \tau \in [\tau_\mu, \tau_{\mu+1}).
\label{E:InterFlipMonotonicity}
\end{equation}

% 1回目の折り返し時刻以降の偏差の指数減衰性
\tcb{We can now establish the exponential convergence. For any iteration $\tau$, let $\mu(\tau)$ be the unique index such that $\tau \in [\tau_{\mu(\tau)}, \tau_{\mu(\tau)+1})$, representing the most recent flip before or at $\tau$. From \eqref{E:InterFlipMonotonicity}:}
\begin{equation}
|e(\tau)| \leq |e(\tau_{\mu(\tau)})| < \gamma^{\mu(\tau)-1} \eta_0.
\end{equation}

% 1回目の折り返し時刻について
% About the first flip time
\tcc{We also see that the first flip time $\tau_{1}$ is}
\begin{equation}
    \lceil |e(0)|/\eta_0 \rceil =: D.
\label{E:FirstFlipTime}
\end{equation}
%% 具体的な減衰率
\tcc{Since the first flip time is $D$ from \eqref{E:FirstFlipTime} and each flip interval is bounded by $C$ from \eqref{E:FlipIntervalBound}, the total number of flips by iteration $\tau$ satisfies $\mu(\tau) \geq D + \lfloor (\tau-D)/C \rfloor$. Therefore:}
\begin{equation}
|e(\tau)| < \gamma^{D + \lfloor (\tau-D)/C \rfloor - 1} \eta_0 \leq \gamma^{D + (\tau-D)/C - 2} \eta_0 = \alpha \gamma^{\tau/C},
\end{equation}
\tcc{where $\alpha = \gamma^{D-D/C-2} \eta_0$.}

% パラメータの次元が複数だった時の一般論
\tcb{For the general vector case $q > 1$, the parameter update rule \eqref{E:M.ParameterUpdater} operates element-wise, so each component $\theta_i(\tau)$ evolves independently according to the scalar dynamics analyzed above. Taking the Euclidean norm:}
\begin{equation}
\|\theta(\tau) - \theta^*\|_2 \leq \sqrt{\sum_{i=1}^q \alpha_i^2} \gamma^{\tau/C} = \alpha \gamma^{\tau/C},
\end{equation}
\tcb{where $\alpha = \sqrt{\sum_{i=1}^q \alpha_i^2}$ with appropriate component-specific constants $\alpha_i$, completing the proof.}

\end{proof}

\begin{figure}
    \centering
    \includegraphics[width=1\linewidth]{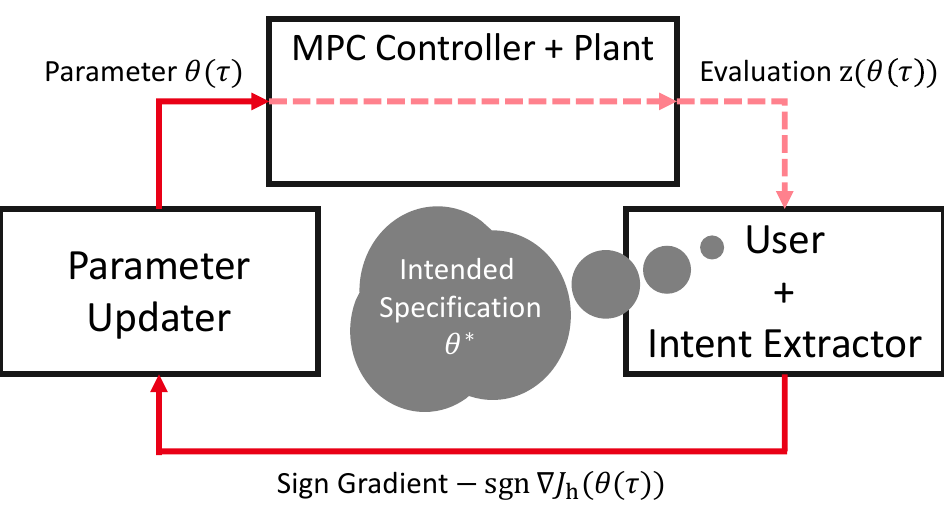}
    \caption{Reformulation of personalization loop \tcb{into} three blocks: the parameter updater standalone, the combination of the MPC controller and the plant, and the combination of the user and the intent extractor.}
\label{F:T.PersonalizationLoop}
\end{figure}

% 有限時間内の収束の証明
\subsection{\tcb{Finite-Time Convergence Analysis}}\label{Subsection:FiniteTimeAnalysis}

\tcb{This analysis considers that the user has an acceptance region around their preferred parameter values, leading to finite-time convergence guarantees that better reflect practical user behavior.}

\begin{assumption}[User Acceptance Region]\label{A:AcceptanceRegion}
\tcb{The user has an acceptance region $\mathcal{A} = \prod_{i=1}^q [\theta^*_i - \epsilon_i, \theta^*_i + \epsilon_i] \subset \mathbb{R}^q$ around their ideal parameter $\theta^*$, where $\epsilon_i > 0$ represents the tolerance for the $i$-th parameter component. The user provides corrective feedback through prompt $p(\tau)$ if and only if $\theta(\tau) \notin \mathcal{A}$, and remains silent when $\theta(\tau) \in \mathcal{A}$.}
\end{assumption}

\begin{theorem}[Finite-Time Convergence]\label{T:FiniteTimeConvergence}
\tcb{Under Assumption~\ref{A:AcceptanceRegion}, the ChatMPC personalization loop achieves finite-time convergence to the user acceptance region. Specifically, there exists a finite time $\tau^* < \infty$ such that $\theta(\tau^*) \in \mathcal{A}$ and no further user interaction is required for $\tau \geq \tau^*$.}

\tcb{Moreover, the convergence time satisfies:}
\begin{equation}
\tau^* \leq \tcc{\max_{i=1,\ldots,q}} \left\lceil \log_\gamma\left(\frac{\eta_{0,i}}{2\epsilon_i}\right) \right\rceil \cdot \lceil \gamma^{-1} \rceil,
\label{E:FiniteTimeConvergenceBound}
\end{equation}
\tcb{where $\eta_{0,i}$ is the initial update size for the $i$-th parameter component.}
\end{theorem}

\begin{proof}
\tcb{The key insight is that contradictory user feedback effectively performs bisection on the parameter space. Since the parameter update rule operates element-wise, we analyze each component $\theta_i$ independently.}

\tcb{For the $i$-th component, consider the scenario where the user provides contradictory feedback at two different iterations. Suppose at iteration $\tau_1$ the user indicates that $\theta_i(\tau_1)$ is "too large" (resulting in $s_i(\tau_1) = -1$), and at iteration $\tau_2 > \tau_1$ indicates that $\theta_i(\tau_2)$ is "too small" (resulting in $s_i(\tau_2) = +1$). Under Assumption~\ref{A:AcceptanceRegion}, this implies:}
\begin{equation}
\theta_i(\tau_1) > \theta^*_i + \epsilon_i \quad \text{and} \quad \theta_i(\tau_2) < \theta^*_i - \epsilon_i.
\end{equation}

\tcb{Consequently, the optimal parameter $\theta^*_i$ must lie within the interval $[\theta_i(\tau_2), \theta_i(\tau_1)]$, and the search space has been effectively contracted to this interval.}

\tcb{The width of this contracted interval is determined by the parameter evolution between $\tau_1$ and $\tau_2$. If $m$ flips occurred for component $i$ between these iterations, then from the exponential analysis in Theorem~\ref{T:ExponentialConvergence}:}
\begin{equation}
|\theta_i(\tau_1) - \theta_i(\tau_2)| \leq 2\gamma^{m-1} \eta_{0,i}.
\end{equation}

\tcb{The search terminates when this interval width becomes smaller than the acceptance region width $2\epsilon_i$:}
\begin{equation}
2\gamma^{m-1} \eta_{0,i} < 2\epsilon_i \quad \Rightarrow \quad m > \log_\gamma\left(\frac{\eta_{0,i}}{2\epsilon_i}\right).
\end{equation}

\tcb{Therefore, component $i$ requires at most $\lceil \log_\gamma(\eta_{0,i}/(2\epsilon_i)) \rceil$ flips to enter its acceptance region.}

\tcb{From the exponential analysis, each flip interval is bounded by $\lceil \gamma^{-1} \rceil$. Hence, the iterations required for component $i$ to converge satisfies:}
\begin{equation}
\tau^*_i \leq \left\lceil \log_\gamma\left(\frac{\eta_{0,i}}{2\epsilon_i}\right) \right\rceil \cdot \lceil \gamma^{-1} \rceil.
\end{equation}

\tcb{Since all components must satisfy their respective tolerance requirements for the system to reach the acceptance region $\mathcal{A}$, the overall convergence time is:}
\begin{equation}
\tau^* = \max_{i=1,\ldots,q} \tau^*_i \leq \sum_{i=1}^q \tau^*_i,
\end{equation}
\tcb{establishing \eqref{E:FiniteTimeConvergenceBound}.}
\end{proof}

\begin{corollary}[Logarithmic User Interaction Complexity]\label{C:InteractionComplexity}
\tcb{The total number of user prompts required for convergence to the acceptance region scales logarithmically with the precision requirements:}
\begin{equation}
\text{Total prompts} = O\left(\tcc{\max_{i=1,\ldots,q}} \log_\gamma\left(\frac{\eta_{0,i}}{\epsilon_i}\right)\right).
\end{equation}
\tcb{This logarithmic scaling demonstrates that the cognitive burden on the user remains manageable even for stringent tolerance requirements.}
\end{corollary}
% < Main/TheoreticalAnalysis.tex <
% > Main/ExperimentCBFMPC.tex >
\section{\tcb{Experimental Validation}}\label{Section:ExperimentalValidation}

% 前振り（実験全体）
\tcb{
This section presents two comprehensive experiments that validate ChatMPC across different control domains and personalization scenarios. 
The first experiment demonstrates personalization of control system parameters through a double integrator robot navigation task, where users can adjust obstacle avoidance behavior via natural language feedback. 
The second experiment showcases real-time specification reconfiguration in a semi-autonomous driving scenario, where human drivers assist the control system by reporting unrecognized obstacles through conversational interaction. }

\subsection{\tcb{Personalization in Academic Navigation Task}}\label{Section:ExperimentCBFLLM}
% この実験の概要

\begin{figure} [t]
    \centering
\includegraphics[width=1\linewidth]{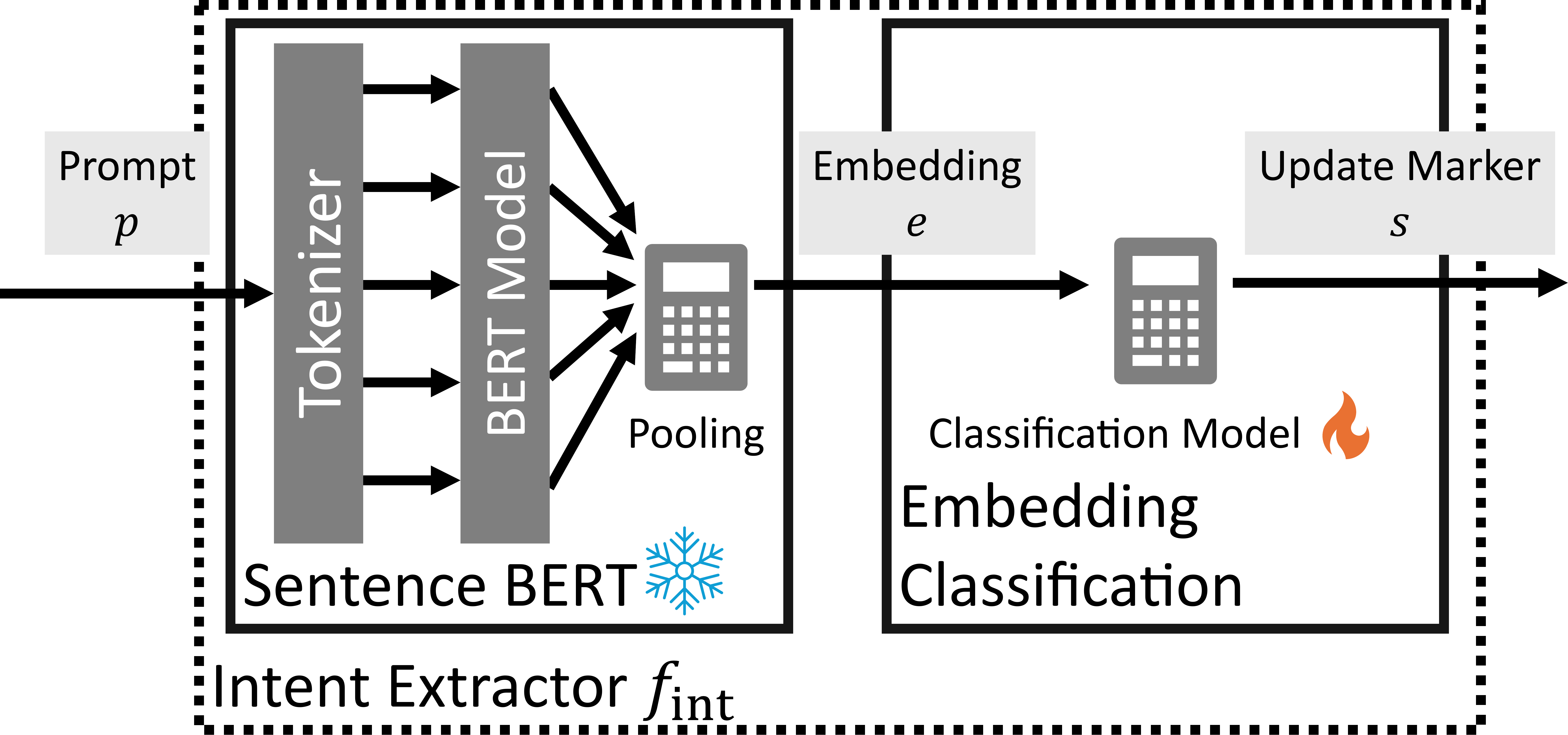}
\caption{Structure of the intent extractor: 
    Sentence BERT consists of a tokenizer, a BERT model, and a pooling. The tokenizer tokenized the provided prompt $p$ into some tokens.
    The tokens are processed by the BERT model, and the BERT model outputs the vectors.
    The pooling calculates the average the vectors and outputs the embedding $e$.}
\label{F:CBFMPC.IntentExtractor}
\end{figure}

% 前振り
\tcb{
This experiment demonstrates ChatMPC on a double integrator robot navigating through obstacles. The robot state $x:=[x_1~x_2~v_1~v_2]^\top \in\mathbb{R}^4$ represents position $(x_1, x_2)$ and velocity $(v_1, v_2)$ components, with discrete-time double integrator dynamics:}
\begin{align}
    x(k+1) =
    \begin{bmatrix}
        1&0&\Delta t&0\\
        0&1&0&\Delta t\\
        0&0&1&0\\
        0&0&0&1
    \end{bmatrix}
    x(k)+
    \begin{bmatrix}
        0&0 \\
        0&0 \\
        \Delta t&0\\
        0&\Delta t
    \end{bmatrix}
    u(k),
    \label{E:CBFMPC.Plant}
\end{align}
\tcb{where $u\in\mathbb{R}^2$ is the control input and $\Delta t = 0.2$ s.}

% 制御目標
\tcb{
The robot must reach a goal while avoiding a vase and a toy, with avoidance behavior personalized through natural language interaction. 
To this end, the robot starts from the initial state $x_0 \neq (0,0)$ and must 
navigate  to the goal position $(0,0)$ while avoiding these obstacles utilizing the MPC structure presented in \eqref{E:M.MPCOptimizationProb}, with  cost function:}
\begin{align}\begin{split}
    & J(X(k), U(k)) :=\\
    & \left[ \sum_{i=0}^{H-1} l(x(k+i+1|k),u(k+i|k)) \right]
    + \phi(x(k+H|k)),
\end{split}\end{align}
\tcb{where the stage cost $l(x,u) = x^\top \text{diag}(1,1,1,1) x + u^\top \text{diag}(1,1)u$ and terminal cost $\phi(x) =x^\top \text{diag}(10^3,10^3,10^3,10^3) x$ drive the robot toward the goal.}

% 障害物回避のためのCBFの導入
\tcb{
For each obstacle $o\in\{1,2\}$, we define the control barrier function (CBF) \cite{Ames19, Zeng21}:}
\footnote{\tcb{A function $h$ is a CBF if there exists a control input $u$ such that:}
\begin{equation}
    \dot h(x) = \frac{\partial h(x)}{\partial x}f(x,u) \geq -\beta(h(x)),
\label{E:MPCCBF.CBFFullVersion}
\end{equation}
\tcb{where $f$ characterizes the plant dynamics $\dot x=f(x,u)$, and $\beta(h)$ is a continuously differentiable strictly increasing function with $\beta(0)=0$. 
For discrete-time MPC implementation, we convert this into:}
\begin{equation}
    \Delta h_i \geq -\beta_{\rm d} h(x(k+i|k)),
\label{E:MPCCBF.CBFEasyVersion}
\end{equation}
\tcb{where  $\Delta h_i=h_{i+1}-h_i$ and $\beta_{\rm d}\in(0,1]$ is a constant. 
This constraint ensures forward invariance: if $h(x(0))\geq0$ initially and \eqref{E:MPCCBF.CBFEasyVersion} holds for all times, then $h(x(k))\geq0$ is maintained.}}% 備考終わり
\begin{align}
    h_o(x) = (x_1-x_{o,1})^2 + (x_2-x_{o,2})^2 - R_o^2,
    \label{E:CBFMPC.CBFCandidate}
\end{align}
\tcb{which keeps the robot outside the safety margin $R_o$ of obstacle $o$ located at $(x_{o,1}, x_{o,2})$ and is synthesized through the CBF constraint:}
\begin{align}
    \Delta h_{o,i} \geq - \theta_o h_{o,i},
\label{E:CBFMPC.CBFInequality}
\end{align}
\tcb{where $\theta_o$ is the personalization parameter that can be modified via ChatMPC. 
In this way, the user can control the conservativeness of avoidance behavior, since higher values of $\theta_o$ result in more cautious behavior around obstacle $o$. 
In our setup, obstacle $o=1$ is the vase and obstacle $o=2$ is the toy, giving us the personalization vector $\theta = [\theta_1 ~ \theta_2]^\top$ that users can adjust through natural language feedback.}

% 意図抽出器
\tcb{The intent extractor $\IntentExtractor$ employed to convert natural language prompts into parameter updates\footnote{The update magnitude is controlled by the constant $d=[2~2]^\top$ in the parameter updater \eqref{E:M.ParameterUpdater}.} is represented in Figure~\ref{F:CBFMPC.IntentExtractor} and uses a two-stage approach:}

\begin{enumerate}
    \item \tcb{First, Sentence BERT with pre-trained weights from ``deepset/sentence\_bert'' \cite{Deepset21} converts user's prompt $p$ into contextual embeddings $e\in\mathbb{R}^D$, which represent the  latent information of the sentence \cite{Nils19}. The model consists of a tokenizer, BERT encoder, and pooling layer that outputs fixed-size representations regardless of input length.}
    \item \tcb{Second, a k-nearest neighbors classifier maps the embedding $e$ to one of four intent classes $s$, each associated with a specific parameter update direction. Some example prompts and update markers for the training data of the classification model are shown in Table~\ref{T:CBFMPC.TrainData}.}
\end{enumerate}

{\small
\begin{table}[t]
    \centering
    \caption{Prompts and update markers for the training data}
    \begin{tabular}{l|c}
        \hline
        Example prompts & Update marker  \\
         \hline

         % SV
         \begin{tabular}{l}
            ``Can you separate from the vase?''\\
            ``Please separate from the vase.''\\
            ``It is too close to the vase.''\\
            ``Too close to the vase.''\\
            ``You are too close to the vase''\\
        \end{tabular}
        & $[-1~0]^\top$ \\
        \hline

        % AV
        \begin{tabular}{l}
            ``Can you approach to the vase?''\\
            ``Please approach to the vase.''\\
            ``You do not need to care about the vase.''\\
            ``You do not need to be careful about the vase.''\\
            ``You do not have to care about the vase so much.''\\
        \end{tabular}
        & $[+1~0]^\top$ \\
        \hline

        % ST
        \begin{tabular}{l}
            ``Can you separate from the toy?''\\
            ``Please separate from the toy.''\\
            ``It is too close to the toy.''\\
            ``Too close to the toy.''\\
            ``You are too close to the toy''\\
        \end{tabular}
        & $[0~-1]^\top$ \\
        \hline

        % AT
        \begin{tabular}{l}
            ``Can you approach to the toy?''\\
            ``Please approach to the toy.''\\
            ``You do not need to care about the toy.''\\
            ``You do not need to be careful about the toy.''\\
            ``You do not have to care about the toy so much.''\\
        \end{tabular}
        & $[0~+1]^\top$ \\
         \hline
    \end{tabular}
    \label{T:CBFMPC.TrainData}
\end{table}
}

\begin{figure*}[t] % * para abarcar dos columnas en documentos tipo IEEEtran
    \centering
    % Imagen A
    \begin{subfigure}{0.49\linewidth}
        \centering
        \includegraphics[width=\linewidth]{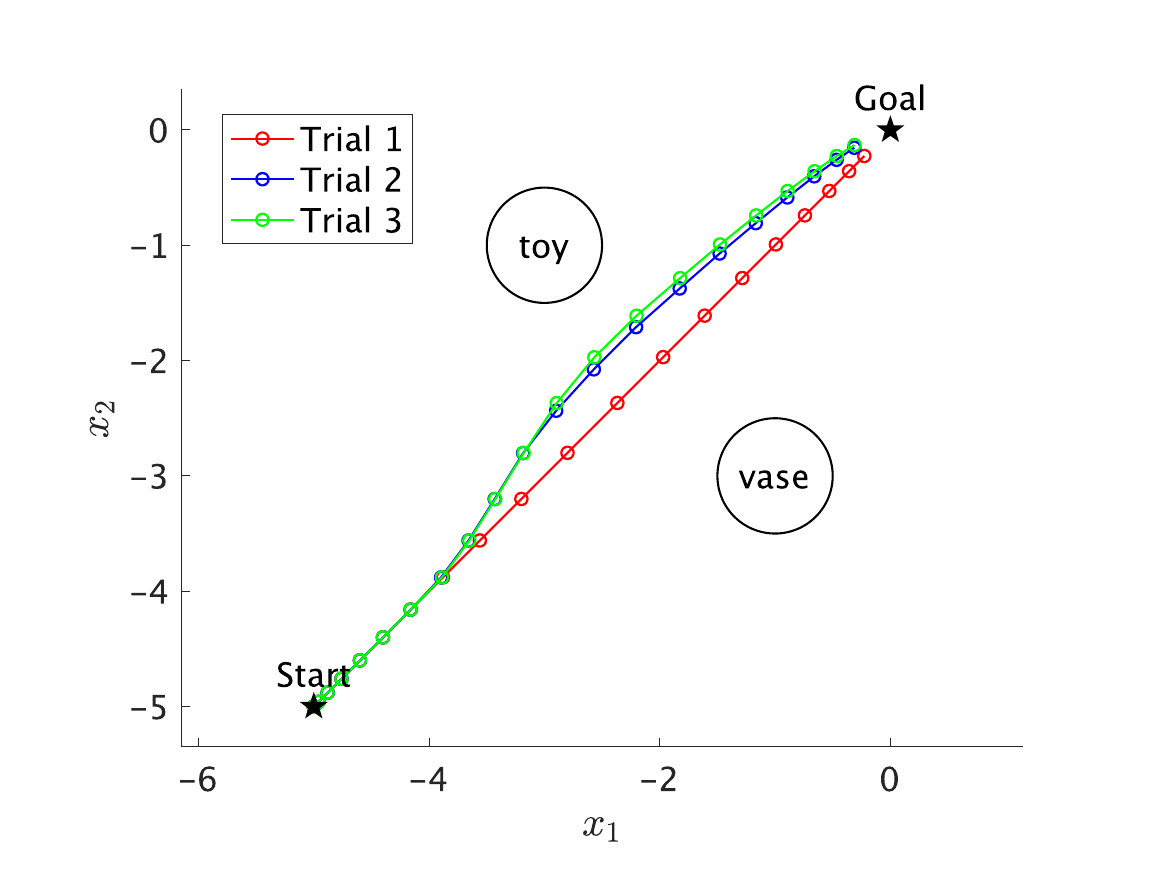}
        \caption{Environment A}
        \label{F:CBFMPC.Result.EnvA}
    \end{subfigure}
    \hfill
    % Imagen B
    \begin{subfigure}{0.49\linewidth}
        \centering
        \includegraphics[width=\linewidth]{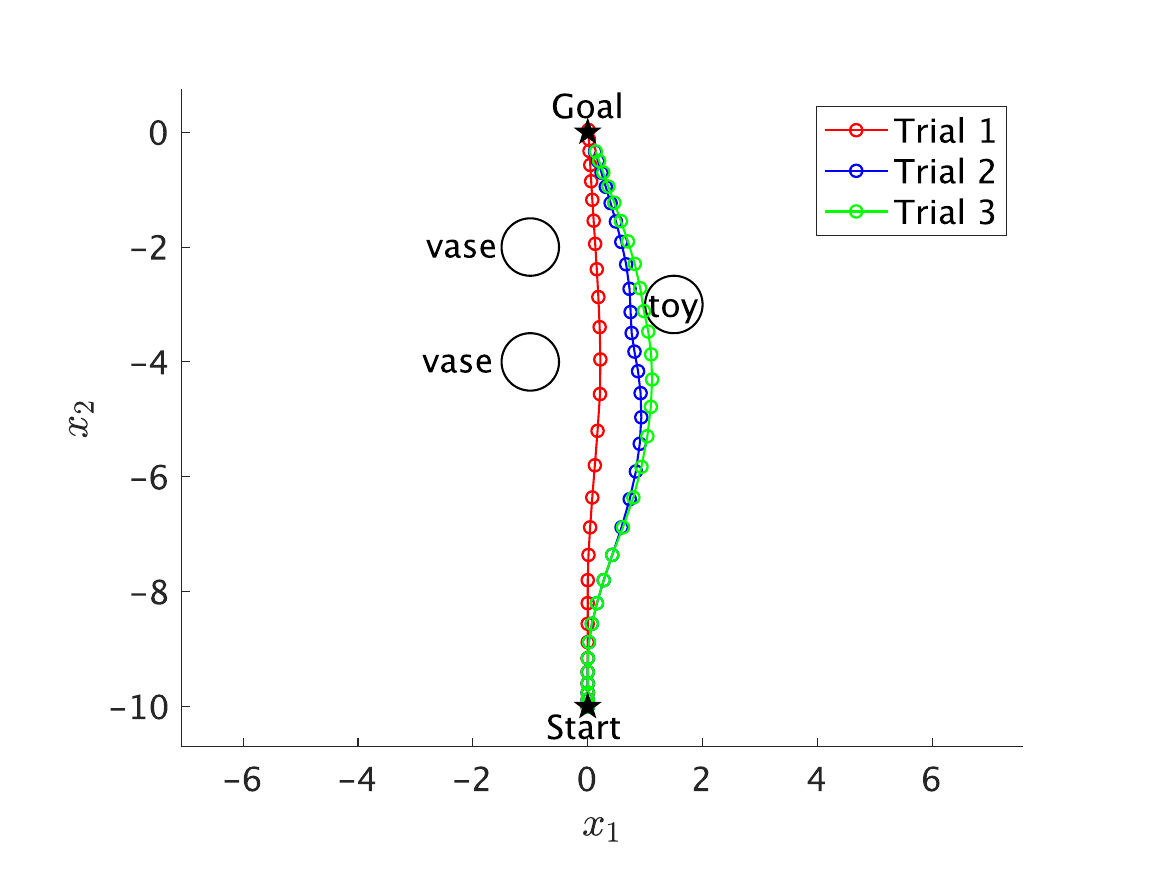}
        \caption{Environment B}
        \label{F:CBFMPC.Result.EnvB}
    \end{subfigure}

    \caption{Robot trajectory in each trial for two different environments.}
    \label{F:CBFMPC.Result}
\end{figure*}

\tcb{To assess the consistency of personalization of ChatMPC, we consider two environments with different obstacle configurations:}
\begin{enumerate}[label=\Alph*)]
    \item \tcb{ The vase is positioned at $(-1,-3)$, the toy at $(-3,-1)$, and the robot starts at $x(0)=[-5~-5~0~0]^\top$.}
    \item \tcb{The vase is positioned at $(-1,-4)$, the toy at $(1.5,-3)$, and the robot starts at $x(0)=[0~-10~0~0]^\top$.}
\end{enumerate}

\tcb{Both environments use safety margin $R=0.5$ for all obstacles and initial parameters $\theta_0=[0.4~0.4]^\top$. Each environment involves three sequential trials with progressive personalization:}

\begin{enumerate}
    \item No prompt: \tcb{Baseline behavior without user feedback}.
    \item \tcb{User prompt: ``Separate from the vase.''}
    \item \tcb{User prompt: ``You don't have to be so careful about the toy.''}
\end{enumerate}

The robot's trajectories in environment A are shown in Figure~\ref{F:CBFMPC.Result.EnvA}, and \tcb{those} in environment B are shown in Figure~\ref{F:CBFMPC.Result.EnvB}. \tcb{With each trial, the robot's trajectory progressively adapts to user preferences, moving away from the vase obstacle (Trial 2) and toward the toy obstacle (Trial 3) as requested through natural language feedback. Note that after the  prompt in Trial 3, the robot adopts a less conservative approach toward the toy while preserving the increased vase avoidance from Trial 2, showing successful accumulation of multiple user preferences. The consistent behavior changes observed across both environments, despite different obstacle layouts and starting positions, indicate that ChatMPC learns generalizable user preferences, confirming the robustness of the personalization mechanism.}

% < Main/ExperimentCBFMPC.tex <
% > Main/ExperimentCarla.tex >
\subsection{Task Co-Development for Autonomous Vehicle in CARLA}\label{Section:ExperimentCarla}
% 前振り
\tcb{This experiment considers a semi-autonomous driving scenario where real-time specification reconfiguration occurs via ChatMPC. In particular, while the vehicle's control system handles most driving tasks —path planning, tracking, and collision avoidance—, the human driver provides safety oversight that can complement the imperfect information from the vehicle's obstacle sensors, which may fail to recognize obstacles that are clearly visible to the human driver. }

\begin{figure}[t]
    \centering
    % Primera imagen
    \begin{subfigure}{\linewidth}
        \centering
        \includegraphics[width=0.9\linewidth]{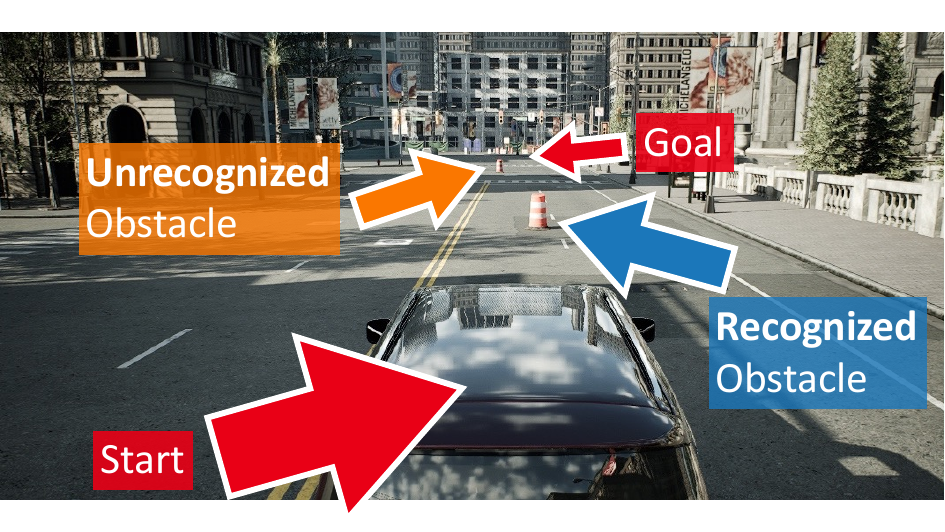}
        \caption*{(a)}
    \end{subfigure}
    
    % Segunda imagen
    \begin{subfigure}{\linewidth}
        \centering
        \includegraphics[width=0.9\linewidth]{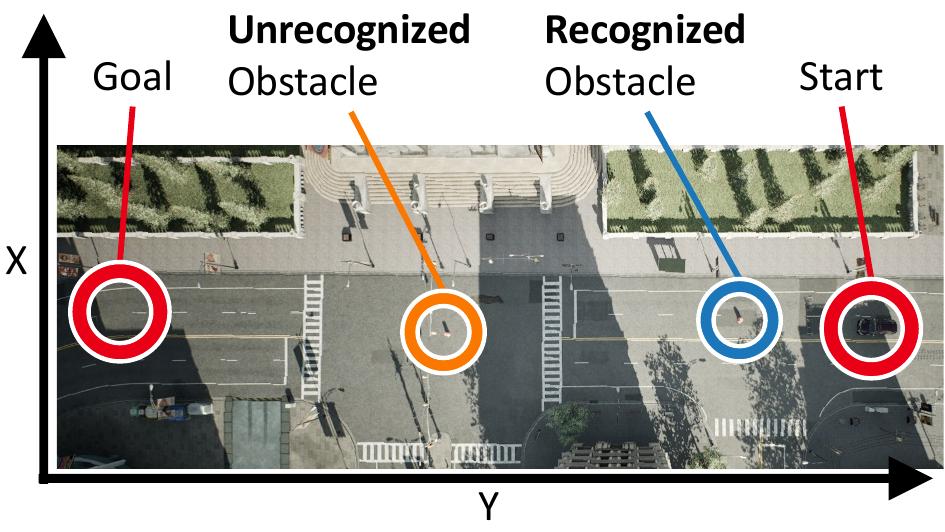}
        \caption*{(b)}
    \end{subfigure}
    
    \caption{Simulation environment: (a) driver view, (b) bird view}
    \label{F:Carla.EnvironmentOverview}
\end{figure}

% シミュレータの紹介
\tcb{We implement this scenario using CARLA simulator \cite{Carla} version 0.10.0 with a defined start point, goal point, and two obstacles positioned between them. The vehicle recognizes one obstacle through its sensors but fails to detect the other, creating a realistic sensor limitation scenario. Figure~\ref{F:Carla.EnvironmentOverview} shows the simulation environment from the driver's perspective and bird's eye view, respectively.}

% プラント
\tcb{The controlled vehicle is a Nissan Patrol with default CARLA physics}\footnote{\tcb{The Blueprint ID of this vehicle is ``vehicle.nissan.patrol'' in CARLA.}} \tcb{whose dynamics are approximated using a bicycle model. Letting $z=[x~y]^\top$ and $\phi$ denote the vehicle position and orientation respectively, the discrete-time dynamics are described by:}
\begin{subequations}
\begin{empheq}[left={\empheqlbrace}]{align}
    z(k+1) &= z(k) + v(k) \Delta t 
    \begin{bmatrix}
        \cos \phi(k) \\ \sin \phi(k)
    \end{bmatrix},\\
    \phi(k+1) &= \phi(k) + \frac{v(k)}{L} \Delta t \delta(k),
\end{empheq}
\end{subequations}
\tcb{where $k$ is the discrete time step with interval $\Delta t=0.04$ s, $v(k)$ is the vehicle speed controlled by a proportional integral controller, $\delta(k)$ is the steering angle serving as our control input, and $L=13.90$ m is the wheelbase determined through system identification applied to driving data.}

% 制御器
\tcb{To solve the MPC optimization problem efficiently, we apply Model Predictive Path Integral (MPPI) control}\footnote{\tcb{MPPI generates $N$ candidate control sequences by sampling from a bounded normal distribution: $U_n \sim \mathcal{N}_{[\delta_{\text{min}},\delta_{\text{max}}]}(0, \sigma^2 I_H)$, where each sample is bounded by steering limits. The optimal control sequence is computed as a weighted average: $$U^* = \frac{\sum_{n=1}^N w_n U_n}{\sum_{n=1}^N w_n}$$ with weights $w_n = \exp(-J(U_n)/T)$ based on the cost function evaluation.}} \cite{Asmar23_MPPI}\cite{Graby17_MPPI} \tcb{with parameters set as sample size $N=500$, prediction horizon $H=50$, steering angle bounds $[-1.0, +1.0]$ rad, standard deviation $\sigma=0.5$, and temperature parameter $T=1.0$.}

% 制御目的
\tcb{The objectives of the MPC controller are: (i) goal-reaching towards a region $\mathcal{S}_{\text{goal}}$, and (ii) obstacle avoidance regarding detected and virtual obstacles, denoted respectively as $S_r$ and $S_v(\theta)$. To these ends, the cost function penalizes the predicted trajectory $Z_{1:H} := [z(1) \cdots z(H)]$ as:}
\begin{align}
    J(Z_{1:H}) = J_{\text{goal}}(Z_{1:H}) + w_{\text{obs}} J_{\text{obs}}(Z_{1:H}),
\end{align}
\tcb{with $w_{\text{obs}}=1000$ and}
\begin{align}
    J_{\text{goal}}(Z_{k:H}) = \begin{cases}
        0 & \text{if } z(k) \in \mathcal{S}_{\text{goal}} \\
        \text{dist}(z(k))^2 + J_{\text{goal}}(Z_{k+1:H}) & \text{otherwise}
    \end{cases}
\end{align}
\begin{align}
    J_{\text{obs}}(Z_{1:H}) = \sum_{k=1}^H \rho^k I(z(k)), \quad I(z) = \begin{cases}
        1 & \text{if } z \in S_r \cup S_v(\theta) \\
        0 & \text{otherwise}
    \end{cases}
\end{align}
\tcb{As can be seen, $J_{\text{goal}}$ is designed recursively to ensure smooth approach towards the goal region $\mathcal{S}_{\text{goal}}$. As for $J_{\text{obs}}$, the decaying rate based on the weight $\rho \in (0,1]$ and the indicator function $I(z)$ penalize more heavily earlier trajectory points that intersect with either recognized or virtual obstacles.}

\begin{figure}
    \centering
    \includegraphics[width=1\linewidth]{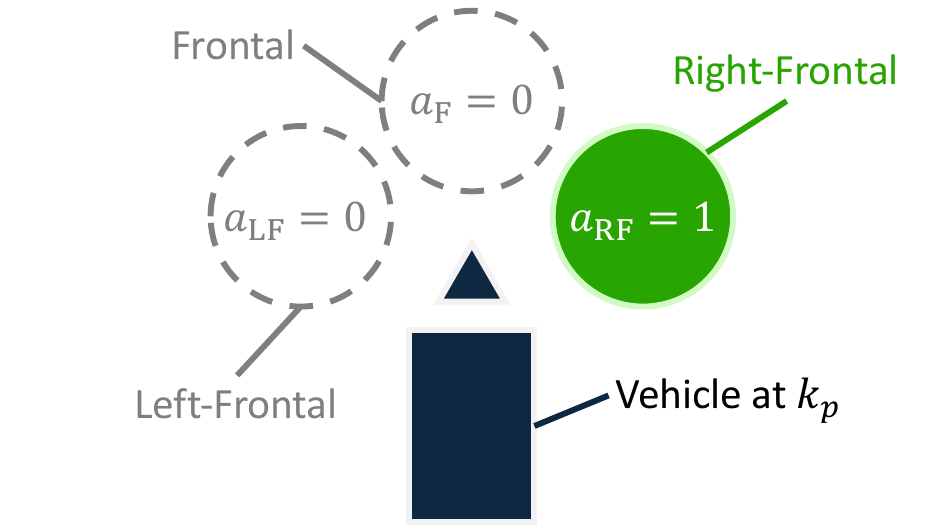}
    \caption{The virtual obstacle $S_\mathrm{v}$ when $a_\mathrm{RF}$ gets 1 at $k_p$}
    \label{F:Carla.VirtualObstacleParametrization}
\end{figure}

\tcb{The virtual obstacle generation mechanism constitutes the core of the personalization approach and follows the previously presented architecture where Sentence BERT extracts embeddings that are classified using a k-nearest neighbors approach. In this case, the intent extractor classifies driver prompts into three spatial categories corresponding to undetected obstacle locations, namely left-frontal (LF), frontal (F), and right-frontal (RF), as illustrated in  the training dataset of Table~\ref{T:Carla.TrainData}, which yield a parameter $\theta = [a_{\text{LF}} ~ a_{\text{F}} ~ a_{\text{RF}}]^\top \in \mathbb{R}_{\geq0}^3$ establishing the location of the virtual obstacle relative to the vehicle.  In particular, if the driver reports an obstacle at a time instant $k_p$, the virtual obstacle $S_\mathrm{v}(\theta)$ is defined as:}
\begin{align}
    S_v(\theta) = \bigcup_{i \in \{\text{LF}, \text{F}, \text{RF}\}} \{ z \mid \|z - z_i(k_p)\|_2 \leq r \cdot a_i \},
\end{align}
\tcb{where $r=4$ m is the baseline safety margin and $z_i(k_p)$ is calculated relative to the vehicle's current position and orientation:}
\begin{subequations}
\begin{empheq}[left={\empheqlbrace}]{align*}
    z_\mathrm{LF}(k) &= z(k) +10 \ [\cos(\phi(k)-0.2) ~ \sin(\phi(k)-0.2)]^\top , \\
    z_\mathrm{ F}(k) &= z(k) +20 \ [\cos\phi(k) \; \sin\phi(k)]^\top , \\
    z_\mathrm{RF}(k) &= z(k) +10 \ [\cos(\phi(k)+0.2) ~ \sin(\phi(k)+0.2)]^\top .
\end{empheq}
\end{subequations}
\tcb{Note that, as before, the the system updates the parameter through $\theta(\tau+1) = \theta(\tau) + s(\tau)$, where $s(\tau) \in \{[1~0~0]^\top, [0~1~0]^\top, [0~0~1]^\top\}$ represents the classified intent. For example, if the driver says \emph{There is an obstacle on the right front!}, then only the right-frontal parameter is activated with $\theta=[0~0~1]^\top$ at step $k_p$, so that a circular virtual obstacle with radius 4 m appears in the right-frontal position relative to the vehicle, as illustrated in Figure~\ref{F:Carla.VirtualObstacleParametrization}.}

{\small
\begin{table}[t]
    \centering
    \caption{Prompts and update markers for the train data}
    \begin{tabular}{l|c}
        \hline
        Example prompts & Update marker  \\
         \hline

         % LF
         \begin{tabular}{l}
            ``There is an obstacle on the left front!''\\
            ``Something is blocking the left front.''\\
            ``Watch out, obstacle ahead on the left!''\\
            ``Obstacle detected to the left front.''\\
            ``Danger on the left front side.''\\
            ``Please avoid the left front.''\\
            ``Left front is not clear.''\\
            ``Be careful, left front is blocked.''
        \end{tabular}
        & $[1~0~0]^\top$ \\
        \hline

        % F
        \begin{tabular}{l}
            ``There is an obstacle right in front!''\\
            ``Obstacle ahead!''\\
            ``Something is blocking the way ahead.''\\
            ``Watch out, obstacle in front.''\\
            ``Obstacle detected in front.''\\
            ``Danger straight ahead.''\\
            ``Please avoid the front.''\\
            ``Front is not clear.''\\
            ``Be careful, front is blocked.''
        \end{tabular}
        & $[0~1~0]^\top$ \\
        \hline

        % RF
        \begin{tabular}{l}
            ``There is an obstacle on the right front!''\\
            ``Something is blocking the right front.''\\
            ``Watch out, obstacle ahead on the right!''\\
            ``Obstacle detected to the right front.''\\
            ``Danger on the right front side.''\\
            ``Please avoid the right front.''\\
            ``Right front is not clear.''\\
            ``Be careful, right front is blocked.''
        \end{tabular}
        & $[0~0~1]^\top$ \\
        \hline

    \end{tabular}
    \label{T:Carla.TrainData}
\end{table}
}

\tcb{To assess the system performance, the experimental setup includes a fixed goal point at $(106.0, -25.0)$, a recognized obstacle $S_r$ at $(107.0, 65.0)$, and an unrecognized obstacle at $(105.0, 22.0)$. Two scenarios with different starting positions and orientations are considered as test cases:}

\begin{enumerate}
    \item \tcb{Start position $(106.0, 85.0)$ with initial direction $-90°$, driver prompt: ``obstacle in front!''}
    \item \tcb{Start position $(104.0, 85.0)$ with initial direction $-120°$, driver prompt: ``an obstacle is present in the right frontal area.''}
\end{enumerate}

\begin{figure*}[t] 
    \centering
    \begin{minipage}{1\linewidth} % ChatMPCCarlaSimulation > AutoDriveTelemetryView
        \centering
        \includegraphics[width=0.7\columnwidth]{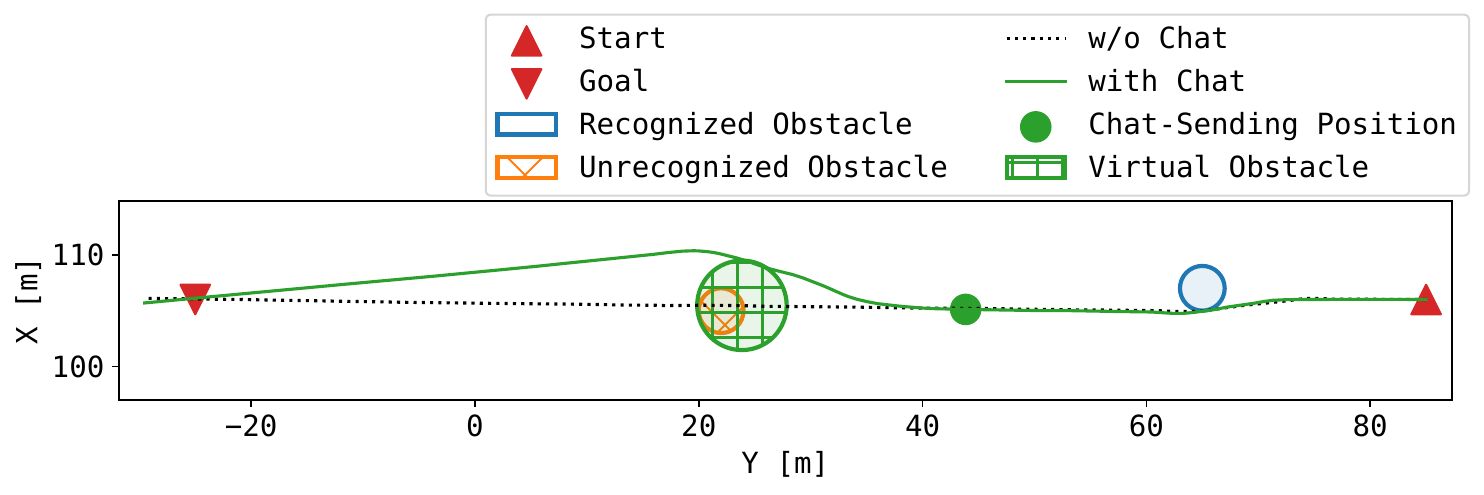}
        \subcaption{Case 1}
        \label{F:Carla.LocationHistories.Scenario1}
    \end{minipage}
    \begin{minipage}{1\linewidth} % ChatMPCCarlaSimulation > AutoDriveTelemetryView
        \centering
        \includegraphics[width=0.7\columnwidth]{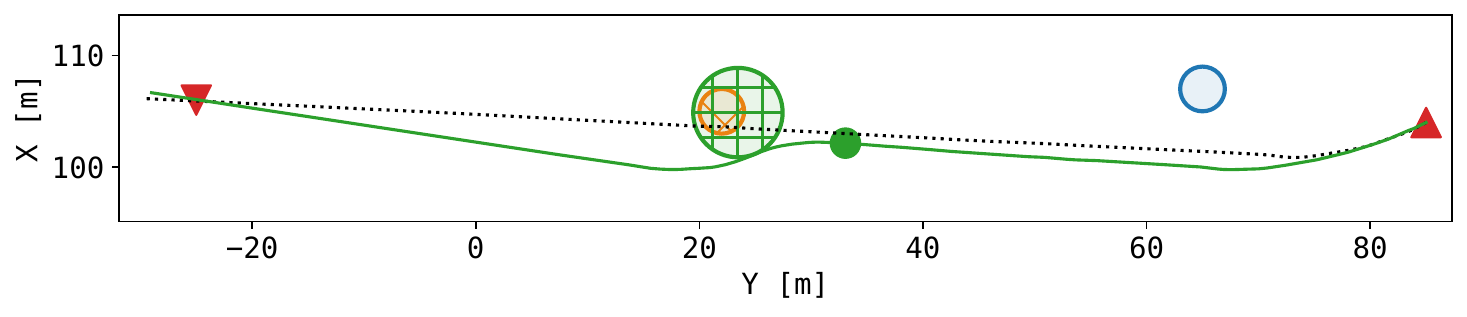}
        \subcaption{Case 2}
        \label{F:Carla.LocationHistories.Scenario2}
    \end{minipage}
    \caption{Trajectories of vehicle. \tcb{In each case, two simulation records are displayed: the dotted black line represents the trajectory without driver input, while the solid green line shows the trajectory when chat assistance is provided. The blue circle indicates the recognized obstacle $S_r$, the orange circle with X-marked hatching represents the unrecognized obstacle, and the green circle with plus-marked hatching shows the generated virtual obstacle $S_v$.}}
    \label{F:Carla.LocationHistories}
\end{figure*}

\tcb{Figure~\ref{F:Carla.LocationHistories} presents the experimental results, showing trajectories for both scenarios with and without driver obstacle reporting. As can be seen,} when no chats regarding obstacles are provided, the vehicle collides with the unrecognized obstacle. In contrast, when the chats regarding obstacles are provided, the vehicle avoids both recognized and unrecognized obstacles to successfully achieve safe driving. \tcb{Hence, the} obstacle reports \tcb{result }in the safe avoidance of the unrecognized obstacle.

\tcb{Finally, the computational performance demonstrates real-time feasibility, with experiments running on an Intel Xeon W-5245 2.98 GHz processor and NVIDIA RTX A5000 GPU. The combined computation time for MPPI and PI controllers averages $3.76 \pm 2.78$ ms per step, considerably lower than the $40$ ms step interval, confirming real-time control capability.}

% < Main/ExperimentCarla.tex <
% > Main/Conclusion.tex >
\section{Conclusion}\label{Section:Conclusion}
% 提案したもの

\tcb{We proposed ChatMPC, a control framework that enables natural language-driven personalization and real-time task co-development in model predictive control systems, allowing users to intuitively modify control behavior and specifications without requiring technical expertise in optimization or control theory. ChatMPC achieves this through a modular architecture consisting of an intent extractor that converts natural language prompts into structured parameter updates, and a parameter updater that safely modifies MPC specifications while maintaining system stability.}

\tcb{Our theoretical analysis established convergence guarantees for the personalization loop by considering users as gradient oracles, showing that practical tolerance-based user behavior leads to bounded adaptation times with logarithmic complexity in precision requirements. These results provide theoretical foundation for the practical deployment of conversational control systems, though further research is needed to address situations where users' behavior deviates from our assumptions, e.g., when they provide contradictory, ambiguous, or potentially unsafe instructions.}

\tcb{The experimental validation across two distinct domains—robot navigation with personalized obstacle avoidance and semi-autonomous driving with real-time obstacle reporting—demonstrates ChatMPC's versatility and effectiveness. However, further research should expand the approach toward other control paradigms and higher-dimensional systems, potentially leveraging more sophisticated language models for enhanced conversational capabilities. Just as AI-aided Kalman filters have enhanced state estimation through machine learning \cite{Revach22_AIAidedKF}\cite{Shlezinger25_AIAidedKF}, ChatMPC represents a step toward language-model-aided control systems that can democratize access to advanced control capabilities through natural conversation.}
% < Main/Conclusion.tex <

\bibliographystyle{ieeetr}
\bibliography{Reference/Experiment, Reference/RelatedResearch}

\end{document}